# Theoretical Formulation of the Phobos, moon of Mars, rate of altitudinal loss.

## 1. Introduction.

It is well established that our Moon is receding from the Earth at the rate of 3.7cm/yr [Yen 1999] . It is also well established that it it is spiraling out until it will get into a geosynchronous orbit [Kaula & Harris 1975]. In this futuristic orbit it will be orbiting in 47 days. But what has not been known commonly that Roche's Limit[Ida et al 1997] of 18,000 Km lies just beyond the inner Geosynchronous Orbit. At the inner Geosynchronous Orbit, length of month = length of day= 5 hours. At that point of accretion from the circumterrestrial impact generated debris, the fully formed Moon experienced a Gravitational Sling Shot Effect [Cook 2005, Dukla et al 2004, Epstein 2005, Jones 2005] which launched it on an outward expanding spiral path. Gravitational Sling Shot is termed as Planet Fly-by-Gravity-Assist maneuver which is routinely used to boost mission spacecrafts to explore the farthest reach of our Solar System. Gravitational sling Shot creates an impulsive torque which gave the orbiting Moon its extra rotational energy with which it continues to spiral out and climb up the potential well created by a much heavier Earth. It was this Gravitational Sling Shot impulsive torque which enabled Charon to spiral out from the inner Geosynchronous Orbit to outer Geosynchronous Orbit where it is in stable equilibrium tidally interlocked with Pluto. If Moon had fallen short of the inner Geosynchronous Orbit it would have been launched on a gravitational runaway inward spiral path to its certain doom. The Phobos is launched on precisely such an inward gravitational runaway collapsing spiral path because it orbits below the synchronous orbit. It is losing its altitude at the rate of 60ft per century which comes out to be 18.29cm/year [www.ozgate.com/infobytes/mars_views.htm]. This paper will arrive at the same rate of altitude loss by applying planetary satellite dynamics.

## 2. Phobos-Mars System.

Phobos and Deimos are the two moons of Mars. They were discovered by Asaph Hall in 1877. The history of the studies of Mars and its moons are given in Table 1.

| Year | Person or Spacecraft | Work done. |
|---|---|---|
| 1659 | Christian Hugens | Drew the first sketch of the dark and bright side of Mars. |
| 1780 | William Herscel | Noted thin Martian Atmosphere. |
| 1877 | Giovanni Schiapaprelli | Drew first detailed map of Martian surface. |
| 1900 | Percival Lowell | Used Lowell Telescope to make drawing of the canals on Martian Surface. |
| 1965 | Mariner 4 | Beamed back 20 photos from first flyby of Mars. |
| 1971 | Mariner 9 | Sent back 7300 images from first ever orbital mission. |
| 1976 | Viking 1 & 2 | First probes to land on Martian Surface and photograph the terrain. |
| July 7,1988 | Phobos 1 | It failed enroute |
| July 12, 1988 | Phobos 2 | It sent 38 high quality images. |
| 1998 | Mars Global Surveyor | It is mapping the whole surface of Mars |
| 2003 | European Space Agency Mars Express | (1) It has revealed the volcanic past of Mars; (2) Icy Promethei Planum , the icy south pole of Mars , has been photographed; (3) In 2008 Atmsphere stripping on Mars and Venus are being simultaneously studied by Mars Express and Venus Express. |

Phobos is the least reflective body in our Solar System largely constituted of carbonaceous chondrite material called Type-C asteroids(lying in outer part of the Asteroid Belt) and captured early in Solar history. Mars Express has revealed that it is relatively red in colour resembling D-Type Asteroids(lying at the outer edge of the main Asteroid Belt). Phobos is thought to be made of ultra primitive material containing carbon as well as ice but it has experienced even less geo-chemical processing than many carbonaceous chondrites. Hence Phobos date of capture is kept at more than 2.5 Gy. We will assume the date of capture at 2.5Gy.

The synchronous orbit of Phobos is 20,400km. Phobos at an orbital radius of 9380km (about 6000km above the Martian surface) and with an orbital period of 7 hrs 39mins is gradually being drawn inward. Altitudinal loss rate is 1.8m/century as given by Wikipedia and 60ft per century as given by www.ozgate.com/infobytes/mars_views.htm It is estimated that within 50 Myr Phobos will crash into Mars[Duxbury 2007]. The New Perspective gives the crash time as 11Myr from now. The altitude loss is at the rate of 20cm per year or 20m per century.

3. **GLOBE-ORBIT PARAMETERS OF PHOBOS & MARS AND THE RELEVANT PARAMETERS OF LOM/LOD.**

Table 2. Globe and Orbit Parameters of the Mars-Phobos system[Chaisson & McMillan 1998, Hannu et al 2003, Moore 2002]

|  | $M$(kg) Mass of Planet | $R_M$ (m) radius of Planet | C (kg-m$^2$) | a (m) | $P_1$ (solar d) | $P_2$ (solar d) | $m$(kg) Mass of moon | M (gm/cc) | m (gm/c.c.) |
|---|---|---|---|---|---|---|---|---|---|
| Mars-Phobos | 6.4191^23 | 3.397^6 | 2.9634^36 | 9.38^6 | 0.319 | 1.026 | 1.1^19 | 3.93 | 2.0 |

C= Moment of Inertia around the spin axis of the Planet= $(2/5)MR_M^2$;
a = semi-major axis of the moon;
$P_1$ = satellite's orbital period;
$P_2$ = planet's spin period;

It has been shown in Basic Mechanics of Planetary Satellites with special emphasis on Earth-Moon System[http://arXiv.org/abs/0805.0100],

$$\Omega/\omega = P_1/P_2 = LOM/LOD = E \cdot a^{3/2} - F \cdot a^2 \tag{1}$$

where $E = J_T/(BC)$;
$F = (m/(1+m/M))(1/C)$;
$B = \sqrt{(GM(1+m/M))} = \sqrt{(G(m+M))}$;
$J_T = C \cdot \omega + (m/(1+m/M))B\sqrt{a}$;
$\omega = (2\pi/P_2) = $ planet's angular spin velocity;
$\Omega = (2\pi/P_1) = $ satellite's angular orbital velocity;

Roots of LOM/LOD=1 give the two geosynchronous orbits $a_{G1}$ and $a_{G2}$.
Using Mathematica the two roots are:
$a_{G1} = 2.04 \times 10^7$ m = 20,400Km;
$a_{G2} = 8.65 \times 10^{18}$ m = $8.65 \times 10^{15}$ Km;

For all practical purposes $a_{G2}$ is infinity and Phobos if it had tumbled beyond $a_{G1}$ it would never evolve out of $a_{G1}$ as already seen in a previous paper [Sharma & Ishwar 2004a,b,c] for very light mass ratio. This same situation exists for the man made satellites around Earth's geo-synnchronous orbit. For man-made satellites there is only one geo-synchronous orbit at 36,000 km above the equator. The other is at infinity.

All the relevant parameters are given in Table 2. As seen from the Table 2 lom/lod by calculation and observation are the same. So we can say that lom/lod equation is correctly derived.

**Table.3. Parameters E, F, $a_{G1}$, $a_{G2}$, $a_R$, lom/lod $_{cal}$ & lom/lod $_{obs}$ of Mars & Phobos.**

|   | B | $J_{spin}$ | $J_{orb}$ | $J_T$ | E | F | $a_{G1}$ (m) | $a_{G2}$ (m) | $a_R$ (m) | $(\omega/\Omega)_{obs}$ | $(\omega/\Omega)_{cal}$ |
|---|---|---|---|---|---|---|---|---|---|---|---|
| M-P | 6.54^6 | 2.1005^32 | 2.1898^26 | 2.1005^32 | 1.083^-11 | 3.683^-21 | 2.04^7 | 8.65^18 | 10.4^6 | 0.3109 | 0.311 |

Here we give the Mathematica Commands used to arrive at Lom/LOD.
First we substitute the numerical values of E and F in LOM/LOD equation:

$E \times x^{1.5} - F \times x^2 /. \{E \to 1.083 \times 10^{-11}, F \to 3.683 \times 10^{-21}\}$

The result is:

$$1.08300000000000018`\!^\wedge\!\text{-}11\, x^{1.5} - 3.68299999999999982`\!^\wedge\!\text{-}21\, x^2 \quad (2)$$

We use this Equation to determine the theoretical value og LOM/LOD. For this we substitute the present value of semi-major axis of Phobos which is $9.38 \times 10^6$ m in Equation 2.

$1.08300000000000018`\!^\wedge\!\text{-}11\, x^{1.5} - 3.68299999999999982`\!^\wedge\!\text{-}21\, x^2 /. x \to 9.38 \times 10^6$

The result is: $0.3111231031746153787`$ which concurs with observed LOM/LOD within observational errors.

We will use Equation 2 to deduce the geo-synchronous orbits and gravity resonance points.
By equating Eq. 2 to Unity and determining the roots we obtain $a_{G1}$ and $a_{G2}$.

Solve[$1.08300000000000018`\!^\wedge\!\text{-}11\, x^{1.5} - 3.68299999999999982`\!^\wedge\!\text{-}21\, x^2 == 1, x$]

$\{\{x \to 2.04290485535965071`\!^\wedge\!7\}, \{x \to 8.64676140767263667`\!^\wedge\!18\}\}$

These are the two values of geosynchronous orbits which have been tabulated in Table 2.

By equating Eq.2 to TWO we obtain $x_2$ the gravity resonance point.

Solve[$1.08300000000000018`\!^\wedge\!\text{-}11\, x^{1.5} - 3.68299999999999982`\!^\wedge\!\text{-}21\, x^2 == 2, x$]

`{{x → 3.24291018020962829`'^7}, {x → 8.64676140767263667`'^18}}`

$x_2 = 3.2429 \times 10^7$ m. At this gravity resonance point Velocity of Recession will be a maxima and hence its derivative will be a zero.

### 4. THE KINEMATICS OF MARS-PHOBOS.

Now we proceed to determine the velocity of recession/approach and setting up of the orbital integral equation to obtain the transit time from the point of capture which we are assumming to be $a_G1$ to the present positon of $9.38 \times 10^6$ m.

From Sharma & Ishwar 2004a, velocity of recession/approach is:
$$da/dt = (K/a^M)(Ea^2 - Fa^{2.5} - \sqrt{a})(2(1+m/M_{mars}))/(mB) \qquad (3)$$

Expressing Eq. 3 in a more conventional coordinate of x,
$$dx/dt = (K/x^M)(Ex^2 - Fx^{2.5} - \sqrt{x})(2(1+m/M_{mars}))/(mB) \qquad (4)$$

From now on x refers to the semi-major axis a of the evolving Satellite. There are two unknowns exponent 'M' and structure constant 'K' in Eq.3. Therefore two unequivocal boundary conditions are required for the complete determination of the Velocity of Recession.

First boundary condition is at $x = x_2$ which is a Gravitational Resonance Point where LOM/LOD = 2 [Rubicam 1975],
i.e. $Ea^{3/2} - Fa^2 = 2$ has a root at $a_2 = x_2$.
At $a_2 = x_2$ the velocity of recession maxima occurs. i.e. $V(x_2) = V_{max}$.

Therefore at $x = x_2$, $(dV(x)/dx)(dx/dt)|_{x_2} = 0$
$\qquad$ (5)

Eq.5. simplifies to the form:
$$E(2-M)x^{1.5} - F(2.5-M)x^2 - (0.5-M) = 0 \text{ where } x = x_2 \qquad (6)$$
From Eq. 6, M (exponent) is determined.

Using Mathematica Command we obtain exponent M:

`(-0.5 + M) / x^0.5 + E × (2 − M) x − F × (2.5 − M) x^1.5 /.`
`{E → 1.083 × 10^−11, F → 3.683 × 10^−21}`

$$\frac{-0.5 + M}{x^{0.5}} + 1.08300000000000018\text{`}^\wedge\text{-}11\,(2 - M)\,x - 3.68299999999999982\text{`}^\wedge\text{-}21\,(2.5 - M)\,x^{1.5}$$

Since the above equation is zero at $x_2$, hence $x_2$ is substituted and the equation equated to ZERO:

$$\frac{-0.5 + M}{x^{0.5}} + 1.08300000000000018\text{`}^\wedge\text{-}11\,(2 - M)\,x - 3.68299999999999982\text{`}^\wedge\text{-}21\,(2.5 - M)\,x^{1.5} \;/.\; x \to 3.24291018020962829\text{`}^\wedge 7$$

$$0.000351207172516702836\,(2 - M) - 6.8014905494453659\text{`}^\wedge\text{-}10\,(2.5 - M) + 0.00017560324618382392\,(-0.5 + M)$$

Solve[0.000351207172516702836` (2 – M) –
  6.8014905494453659`^-10 (2.5` – M) +
  0.00017560324618382392` (–0.5` + M) == 0, M]

{{M → 3.49999806339270591`}}

So we obtain the value of exponent M to be 3.5.

Second boundary condition is at $x = x_1$. This is the point where the sling-shot effect peaks and radial acceleration is at its positive maximum value. Mathematically this is known as the point of inflexion where the second time derivative is zero. If first time derivative is defined as follows :

$$A_{\text{radial}}(X) = (dV(x)/dx)\,(dx/dt) \tag{7}$$

Then the second time derivative of Eq. 7 equated to zero at $x = x_1$ gives the correct value of K (the constant of the Structure Factor).

Therefore $(d\,A_{\text{radial}}(x)/dx)\,(dx/dt)\,|\,x1 = 0$ \hfill (8)

Solution of Eq. 8 gives the correct value of K.

Solution of Eq. 6 gives the value of exponent M and solution of Eq.8 gives the correct value of K. But sling-shot peak point $x_1$ is not known unequivocally therefore we have to arrive at the correct value of K by iteration method. If the Age of the Planet is known we can be sure that the upper limit of the Age of the Satellite has to be equal or less than the Age of Planet. In our case we have assumed Phobos to be captured 2.5Gy ago. The Transit time in making Non-Keplerian journey from $a_{G1}$ to the present position of the Satellite should be equal to 2.5Gy..

Transit time from $a_{G1}$ to present day $a = \int [(1/V(x))\,dx, a_{G1}, a_{present}]$ (9)

Using Eq. 9 and 3 , we can arrive at the correct value of K.
Through several iterations we arrive at K= {{K →2.72108019350420438`^38}}

In Table 4. the various kinematic parameters are given.

**Table 4. Tabulation of Sling-shot point (x1), Gravitational Resonance point (x2) and Structure Factor parameters M, K,**

|  | $X_1$ (m) | Lom/lod$|_{x1}$ | $X_2$ (m) | M | K ($N-m^{M+1}$) |
|---|---|---|---|---|---|
| Phobos | 2.37^7 | 1.25 | 3.24252^7 | 3.49999 | 2.7210802^38 |

## 5. DETERMINATION OF THE KINETICS, TIME CONSTANT OF EVOLUTION, AGE AND EVOLUTION FACTOR.

To determine the time evolution of Phobos orbital path we will have to set up the Velocity of Recession/Approach by substituting the numerical values of different parameters in Equation 4. Following Mathematica Commands are executed for achieving the same:

First the numerical values of E and F are substituted and the velocity expression which is in m/second is converted into m/year by multiplying Eq.4 by $31.5569088 \times 10^6$ seconds per year:

$$(K \div x^M) \times (E \times x^2 - F \times x^{2.5} - \sqrt{x}) \times$$
$$(2 \times (1 + m/M) \div (mB)) \times 31.5569088 \times 10^6 \;/.$$
$$\{E \to 1.083 \times 10^{-11}, F \to 3.683 \times 10^{-21}\}$$

The result is:
$$\frac{1}{mB}\left(6.31138175999999884\text{`}^{\wedge}7\,K\right.$$
$$\left(1 + \frac{m}{M}\right) x^{-M} \left(-\sqrt{x} + 1.08300000000000018\text{`}^{\wedge}{-11}\,x^2 -\right.$$
$$\left.\left. 3.68299999999999982\text{`}^{\wedge}{-21}\,x^{2.5}\right)\right)$$

Next mass of Mars $m_0$ and mass of Phobos $m_1$ are substituted:

$$\frac{1}{m_1 B}\left(6.31138175999999884\text{\textasciigrave}\text{\textasciicircum}7\, K\right.$$
$$\left.\left(1+\frac{m_1}{m_0}\right)x^{-M}\left(-\sqrt{x}+1.08300000000000018\text{\textasciigrave}\text{\textasciicircum}-11\, x^2-\right.\right.$$
$$\left.\left.3.68299999999999982\text{\textasciigrave}\text{\textasciicircum}-21\, x^{2.5}\right)\right) /.$$
$$\{m_1 \to 1.1 \times 10^{19},\ m_0 \to 6.4191 \times 10^{23}\}$$

The result is:
$$\frac{1}{B}\left(5.7377181037287297\text{\textasciigrave}\text{\textasciicircum}-12\right.$$
$$\left. K\, x^{-M}\left(-\sqrt{x}+1.08300000000000018\text{\textasciigrave}\text{\textasciicircum}-11\, x^2-\right.\right.$$
$$\left.\left.3.68300000000000071\text{\textasciigrave}\text{\textasciicircum}-21\, x^{2.5}\right)\right)$$

Next the value of B is substituted:
$$\frac{1}{B}\left(5.7377181037287297\text{\textasciigrave}\text{\textasciicircum}-12\right.$$
$$\left. K\, x^{-M}\left(-\sqrt{x}+1.08300000000000018\text{\textasciigrave}\text{\textasciicircum}-11\, x^2-\right.\right.$$
$$\left.\left.3.68300000000000071\text{\textasciigrave}\text{\textasciicircum}-21\, x^{2.5}\right)\right) /.$$
$$B \to 6.54 \times 10^6$$

The result is:
$$8.77326927175646709\text{\textasciigrave}\text{\textasciicircum}-19$$
$$K\, x^{-M}\left(-\sqrt{x}+1.08300000000000018\text{\textasciigrave}\text{\textasciicircum}-11\, x^2-\right.$$
$$\left.3.68300000000000071\text{\textasciigrave}\text{\textasciicircum}-21\, x^{2.5}\right)$$

Next the value of exponent $M = 3.5$ is substituted:
$$8.77326927175646709\text{\textasciigrave}\text{\textasciicircum}-19$$
$$K\, x^{-M}\left(-\sqrt{x}+1.08300000000000018\text{\textasciigrave}\text{\textasciicircum}-11\, x^2-\right.$$
$$\left.3.68300000000000071\text{\textasciigrave}\text{\textasciicircum}-21\, x^{2.5}\right) /.$$
$$M \to 3.5$$

The result is:
$$\frac{1}{x^{3.5}}\left(8.77326927175646709\text{\textasciigrave}\text{\textasciicircum}-19\right.$$
$$K\left(-\sqrt{x}+1.08300000000000018\text{\textasciigrave}\text{\textasciicircum}-11\, x^2-\right.$$
$$\left.\left.3.68300000000000071\text{\textasciigrave}\text{\textasciicircum}-21\, x^{2.5}\right)\right) \qquad (10)$$

Next this expression is determined at Gravitational Resonance point by substituting $x = x_2$ :

$$\frac{1}{x^{3.5`}} \left( 8.77326927175646709`\text{^-19} \right.$$
$$\left. K \left( -\sqrt{x} + 1.08300000000000018`\text{^-11}\, x^2 - 3.68300000000000071`\text{^-21}\, x^{2.5`} \right) \right) /.$$
$$x \to 3.24291018020962829`\text{^7}$$

The result is:
$$2.57250779183593625`\text{^-41}\, K \tag{11}$$

Through several iterations we have determined that $V_{max}=0.007$ m/year. We equate Equation 11 to 0.007m/yr to obtain the value of K:

Solve[2.57250779183593625`^-41 K == 0.007, K]

The result is:
$$\{\{K \to 2.72108019350420438`\text{^38}\}\}$$

Next the value of K is substituted in Equation10 to obtain the complete expression of Velocity of Recession/Approach:

$$\frac{1}{x^{3.5`}} \left( 8.77326927175646709`\text{^-19} \right.$$
$$\left. K \left( -\sqrt{x} + 1.08300000000000018`\text{^-11}\, x^2 - 3.68300000000000071`\text{^-21}\, x^{2.5`} \right) \right) /.$$
$$K \to 2.72108019350420438`\text{^38}$$

The complete radial Velocity of Recession/Approach result is:
$$\frac{1}{x^{3.5`}}$$
$$\left( 2.38727692476555697`\text{^20} \left( -\sqrt{x} + 1.08300000000000018`\text{^-11}\, x^2 - 3.68300000000000071`\text{^-21}\, x^{2.5`} \right) \right) \tag{12}$$

Equation 12 will give the Radial Velocity at any point in orbital evolution by substituting the corresponding value of 'x'.

By substituting present semi-major axis of Phobos:

$$\frac{1}{x^{3.5}} \left(2.38727692476555697`{\wedge}20 \right.$$
$$\left(-\sqrt{x} + 1.08300000000000018`{\wedge}{-}11\, x^2 - \right.$$
$$\left.\left.3.68300000000000071`{\wedge}{-}21\, x^{2.5}\right)\right) /.$$
$$x \to 9.38 \times 10^6$$

We obtain the present rate of Approach or present rate of Altitudinal Loss:

$-0.199267239722073625`$ m/year

This comes out to be 20cm/year or 20m/century assuming a transit time of 2.3Gy from the point of capture to the present position.
Wikipedia gives 1.8m/century whereas www.ozgate.com/infobytes/mars_views.htm gives 60 ft/century.

## Phobos Crashes to Mars ?

While Deimos orbits at a safe distance from Mars, Phobos is spiraling slowly toward eventual destruction. The planet's gravitational pull is reeling in the moon at a rate of 60 feet per century. But collision between Phobos and Mars may never occur - the moon may suffer the fate of being broken up by the planet's tidal forces.

For determining the transt time from the point of capture to the present position, we will have to solve Equation 9:

$$\text{NIntegrate}\left[\left(1 \div \left(\frac{1}{x^{3.5}} \left(2.38727692476555697`{\wedge}20 \right.\right.\right.\right.$$
$$\left(-\sqrt{x} + 1.08300000000000018`{\wedge}{-}11\, x^2 -$$
$$\left.\left.\left.\left.3.68300000000000071`{\wedge}{-}21\, x^{2.5}\right)\right)\right)\right),$$
$$\{x, 2.04 \times 10^7, 9.38 \times 10^6\}\right]$$

(13)

The Age of Phobos comes to be:
$2.32186966711426157`{\wedge}9$

To determine the time of DOOMSDAY, Equation.13 will have to be determined within the limits of $9.38 \times 10^6$ m to $3.4 \times 10^6$ m :

```
NIntegrate[(1÷(1/x^3.5` (2.38727692476555697`*^20
    (-√x + 1.08300000000000018`*^-11 x^2 −
        3.68300000000000071`*^-21 x^2.5))))‚
  {x, 9.38×10^6, 3.4×10^6}]
```

The result is:
1.03989185912455228`*^7

That is 10.4My from now Phobos will be destroyed.

It is asserted that as soon as Phobos enters 7000km zone above the center of Mars the primary tides will smash it and convert it into annular ring of dust which will eventually spiral into Mars.

Time from now to enter Roche's zone will be:
```
NIntegrate[(1÷(1/x^3.5` (2.38727692476555697`*^20
    (-√x + 1.08300000000000018`*^-11 x^2 −
        3.68300000000000071`*^-21 x^2.5))))‚
  {x, 9.38×10^6, 7×10^6}]
```

7.59569254729103437`*^6

That is in 7.6Myr Phobos will be pulverized into Saturn-like annular ring.

Since Wikipedia and Ozgate are giving conflicting data on rate of altitudinal loss hence actual measurement only can establish if the above analysis is correct.

**Table 5 Tabulation of the kinetics, time constant of evolution $(\tau = (a_{G2} - a_{G1})/V_{max})$, transit time from $a_{G1}$ to $a_{present}$ = Age and the evolution factor $\epsilon = (a-a_{G1})/(a_{G2}-a_{G1})$ of Phobos.**

|  | $V_{max}$ (m/y) | $V_{present}$ (m/y) | $\epsilon$ | $\tau$ | Age |
|---|---|---|---|---|---|
| Phobos | 0.007 | -19.9 | -1.22^-12 | 1.29^19y | 2.3Gy |

**6. CONCLUSION.**

This analysis remains inconclusive as far as the validity of this Gravitational Sling Shot approach is concerned because we donot have an authoritative record of Phobos Altitutdinal rate of loss . Some space mission will have to be aimed at Phobos itself. The space craft equipped with Radar must land on Phobos and carry out Mars Ranging Experiments to get an authoritative record of rate of altitudinal loss.

This experiment could be carried out from Mars itself by sounding Phobos.